\documentstyle[11pt]{article}
\begin{document}
\newcommand{\BE}{\begin{equation}}
\newcommand{\EE}{\end{equation}}
\newcommand{\BA}{\begin{eqnarray}}
\newcommand{\EA}{\end{eqnarray}}
\newcommand{\HAM}{Hamiltonian}
\newcommand{\HD}{Hamiltonian density}
\newcommand{\half}{\frac{1}{2}}
\newcommand{\HL}{\hat{\lambda}}
\newcommand{\LB}{\lambda_B}
\newcommand{\LL}{\lambda}
\newcommand{\LR}{\lambda_R}
\newcommand{\LPH}{\lambda \phi^4}
\newcommand{\ls}{{\cal L}_{{\rm Scalar}}}
\newcommand{\lG}{{\cal L}_{{\rm Gauge }}}
\newcommand{\hs}{{\cal H}_{{\rm Scalar}}}
\newcommand{\hG}{{\cal H}_{{\rm Gauge }}}
\newcommand{\PC}{\varphi_c}
\newcommand{\PRI}{\>'}
\newcommand{\pri}{\>'}
\newcommand{\OC}{\Omega}
\newcommand{\OL}{\omega}
\newcommand{\vx}{\vec x\,}
\newcommand{\vy}{\vec y\,}
\newcommand{\vz}{\vec z\,}
\newcommand{\vp}{\vec p\,}
\newcommand{\del}{\vec \nabla\,}
\newcommand{\delsq}{\nabla^2}
\newcommand{\bm}{|\bar\mu\rangle}
\newcommand{\bu}{\bar\mu}
\newcommand{\TG}{\tilde G}
\newcommand{\TF}{\tilde{\varphi}_c}
\newcommand{\PHA}{\hat{\phi}_1}
\newcommand{\PHB}{\hat{\phi}_2}
\newcommand{\VAC}{| \, 0\!>_{\scriptscriptstyle G}}

\setcounter{page}{1}
{\small July, 1992}
\vspace*{-8mm}
\begin{flushright}
{\small DOE/ER/05096-51}
\vspace{24mm}
\end{flushright}
\begin{center}
{\LARGE\bf Gaussian Effective Potential for the\\
\vspace*{5mm} U(1) Higgs Model}
\vspace{20mm}\\
{\Large R. Iba\~{n}ez-Meier, I. Stancu, and P. M. Stevenson}
\vspace{18mm}\\
{\large\it
T.W. Bonner Laboratory, Physics Department,\\
Rice University, Houston, TX 77251, USA}
\vspace{21mm}\\
\end{center}
{\bf Abstract:}

  In order to investigate the Higgs mechanism nonperturbatively, we compute
the Gaussian effective potential (GEP) of the U(1) Higgs model (``scalar
electrodynamics'').  We show that the same simple result is obtained in
three different formalisms.  A general covariant gauge is used, with Landau
gauge proving to be optimal.  The renormalization generalizes the
``autonomous'' renormalization for $\LPH$ theory and requires a particular
relationship between the bare gauge coupling $e_B$ and the bare scalar
self-coupling $\LB$.  When both couplings are small, then $\lambda$ is
proportional to $e^4$ and the scalar/vector mass-squared ratio is of
order $e^2$, as in the classic 1-loop analysis of Coleman and Weinberg.
However, as $\lambda$ increases, $e$ reaches a maximum value and then
decreases, and in this ``nonperturbative'' regime the Higgs scalar can be
much heavier than the vector boson.  We compare our results to the
autonomously renormalized 1-loop effective potential, finding many
similarities.  The main phenomenological implication is a Higgs mass of
about 2 TeV.

\newpage
\setcounter{footnote}{0}
\setcounter{equation}{0}

\section{Introduction}

   The Higgs mechanism \cite{higgs} is a vital, but problematic, aspect
of the Standard Model.  At the classical level it is clear that spontaneous
symmetry breaking (SSB) in the $\LPH$ scalar sector, through its coupling
to the gauge sector, generates gauge-boson mass terms.  The issue of how
-- or whether -- this works in the full quantum theory can be addressed
using the effective potential \cite{gws}, and traditionally the 1-loop
approximation \cite{cole,jackiw} has been used.  However -- at least as it
is conventionally renormalized -- the 1-loop effective potential (1LEP) is
closely tied to perturbation theory.  The possibility that a perturbative
approach is totally misleading must be raised by the claims that the
$(\LPH)_4$ theory is actually ``trivial'' \cite{aiz}, and by the failure
of lattice Monte-Carlo calculations to find a non-trivial, interacting
theory \cite{call}.  Thus, it is very important to study $(\LPH)_4$ theory
and the Higgs mechanism with nonperturbative methods.

   A simple, nonperturbative method, founded upon intuitive ideas familiar
in ordinary quantum mechanics, is the Gaussian effective potential (GEP)
\cite{barnes,gep1}.  In the appropriate limiting cases it contains the
one-loop and leading-order $1/N$ effective potential results
\cite{barnes,gep1,gep2,ON}.   The GEP for O$(N)$-symmetric $(\LPH)_4$ theory
can be renormalized in {\it two} different ways \cite{dimcont}:  the
``precarious'' renormalization, with a {\it negative} infinitesimal $\LB$
\cite{gepz,gep2,ON}, yields essentially the leading-order $1/N$ result
\cite{cjp}.  The resulting effective potential does not, however,
display SSB.  The other, ``autonomous'', renormalization \cite{auto1,auto2,ON},
which can have SSB, is characterized by a {\it positive} infinitesimal $\LB$
and an infinite re-scaling of the classical field.  The resulting theory is
asymptotically free \cite{asy}, which can explain why the ``triviality''
proofs \cite{aiz} do not apply.  Particle masses turn out to be proportional
to $\langle \phi \rangle$, so in the unbroken-symmetry phase the particles
must be massless.  This can perhaps explain the negative findings of most
lattice calculations \cite{call}.  (See Ref. \cite{kerman} for an interesting
comparison of recent lattice results \cite{polonyi} with the Gaussian
approximation.)  The ``autonomous'' theory cannot be obtained in the $1/N$
expansion because $\LB$ must behave as $1/\sqrt{N}$, not as $1/N$, when
$N \rightarrow \infty$ \cite{ON,largen}.

  We used to believe that the ``autonomous'' theory could only be seen
with the Gaussian (or some still-better) approximation.  However, it has
been shown recently by Consoli and collaborators \cite{consoli}
that the unrenormalized 1LEP can also be renormalized in an
``autonomous''-like way.  This result generalizes to more complicated
theories \cite{ims}.  Applied to the SU(2)$\times$U(1) electroweak theory
it predicts a Higgs mass of about 2 TeV \cite{consoli,ims}.

  In this paper we calculate the GEP for the U(1)-Higgs model, which is
O(2) $\LPH$ theory coupled to a U(1) gauge field.  We show that it
can be renormalized in an ``autonomous''-like fashion, and that the vector
boson aquires a mass proportional to $\langle \phi \rangle$, just as in the
traditional description of the Higgs mechanism.
The bare gauge coupling constant $e_B^2$ and the bare scalar self-coupling
$\LB$, both infinitesimal, are related such that for a given $e_B^2$ (below
some maximum value) there are {\it two} allowed values of $\LB$.  One
of these lies in a ``perturbative'' regime in which $\lambda \sim e^4$,
where the results agree with the classic 1-loop analysis of Coleman and
Weinberg \cite{cole}.  The other lies in a ``nonperturbative'' regime,
where it is possible to have a Higgs particle which is arbitrarily heavy
compared to the vector boson.  (See Figs. 1 and 2.)  Our results have much
in common with the ``autonomously renormalized'' 1LEP \cite{consoli,ims},
and thus tend to support the expectation of a 2 TeV Higgs mass.

  The layout of the paper is as follows:  After some preliminaries in Sect. 2,
we outline three separate calculations of the GEP for the U(1) Higgs
model using three different formalisms: Sect. 3 describes a canonical,
Hamiltonian-based calculation, as in \cite{gep1,gep2}; Sect. 4 gives a
covariant ``$\delta$ expansion'' calculation, as in \cite{paper1} (see also
\cite{oko,delta,gandhi}); and Sect. 5 outlines a covariant variational
calculation, as in \cite{jensen}.  We think it is very instructive, as well
as reassuring, to see how the same result emerges from these very different
approaches.  Some comments on the unrenormalized result are given in Sect. 6,
where we show that the optimal gauge parameter is $\xi=0$ (Landau gauge).
The renormalization of the GEP is carried out in Sect. 7.  To conclude, we
discuss the comparison to the 1LEP results and the implications for the Higgs
mass in Sect. 8.

\section{Preliminaries}
\setcounter{equation}{0}

  We first recall the integrals which play a central r\^{o}le
in the GEP.  The expectation value of $\phi^2$ for a single scalar field
yields the quadratically divergent integral
\BE
I_0(\Omega) = \int \frac{{\rm d}^3 p}{(2 \pi)^3}
\frac{1}{2 \omega_{\underline{p}}},
\quad\quad \omega_{\underline{p}} \equiv \sqrt{\vec{p}^{\,2} + \Omega^2},
\EE
which is equivalent to the contracted Euclidean propagator $G(x,x)$
(``tadpole diagram'') integral
\BE
I_0(\Omega) = \int \frac{{\rm d}^4 p}{(2 \pi)^4}
\frac{1}{p^2 + \Omega^2}
\EE
that arises in the manifestly covariant formalism.  The vacuum energy for
a free scalar field of mass $\Omega$ is given by the quartically divergent
integral
\BE
\label{I1}
I_1(\Omega) = \int \frac{{\rm d}^3 p}{(2 \pi)^3} \; \frac{1}{2}
\omega_{\underline{p}},
\EE
which is the sum of the zero-point energies for each momentum mode.  This
integral is familiar from the 1LEP and in the covariant formalism it arises
in the form
\BE
\frac{1}{2} \mbox{Tr}\: \ln \left[ G^{-1}(x,y) \right] / {\cal V} =
I_1(\Omega)
= \frac{1}{2} \int \frac{{\rm d}^4 p}{(2 \pi)^4}
\ln \left( p^2 + \Omega^2 \right),
\EE
where ${\cal V}$ is the spacetime volume.  (Actually, this form of $I_1$
is only equivalent to the canonical form (\ref{I1}) up to an infinite
constant \cite{gep2}.)  The GEP also naturally involves the combination
\BE
J(\Omega) \equiv I_1(\Omega) - \frac{1}{2} \Omega^2 I_0(\Omega),
\EE
which arises from the expectation value of a {\it massless} scalar-theory
Hamiltonian (i.e., kinetic terms only), evaluated in the vacuum of a free
field theory with mass $\Omega$.

  The GEP is essentially a variational calculation:  one first obtains
a function $V_G$ of the classical field $\PC$ and of the mass parameters,
and then one has to minimize with respect to the mass parameters.  This
leads to coupled ``optimization equations'' for the optimal mass parameters
(denoted by overbars).  In carrying out the minimization one needs the formal
result:
\BE
\frac{{\rm d} I_1(\Omega)}{{\rm d} \Omega^2} = \frac{1}{2} I_0(\Omega).
\EE
Further discussion of these divergent integrals is postponed until Section
7.

  The quantization of gauge theories in a covariant gauge always involves
Faddeev-Popov ghosts.  However, in the U(1) case the ghosts are free.
Since they do not couple to the other fields, they have no effect,
{\it except} for their contribution to the vacuum energy \cite{ambj}.
Because the ghosts correspond to two free, massless, anticommuting degrees
of freedom, their contribution is easily seen to be $-2I_1(0)$.  (In the
covariant formalism this term would come from performing the functional
integral over the ghost fields.)  Since this contribution is $\PC$
independent, it will drop out when the infinite vacuum-energy constant is
subtracted off.  This happens automatically in dimensional regularization,
which effectively sets $I_1(0)=0$.  We shall therefore ignore ghosts in the
following calculations.

%
%

\section{Canonical GEP calculation}
\setcounter{equation}{0}

The Lagrangian for the U(1) Higgs model (ignoring ghosts) is:
\BE
{\cal L} =  \lG + \ls,
\EE
where $\lG$ is discussed below,
and $\ls$ is the Lagrangian for a complex scalar field, $\phi$, with the
derivative replaced by the covariant derivative: \cite{fnte}
\BE
\ls = \left(D_{\mu}\phi\right)^{\ast}
\left(D_{\mu}\phi\right)
- m_B^2\phi^{\ast}\phi - 4\lambda_B \left(\phi^{\ast}\phi\right)^2,
\EE
with
\BE
D_{\mu}=\partial_{\mu} + i e_B A_{\mu},
\EE
where $e_B$ is the bare gauge coupling constant.  Replacing the complex
field by two real fields:
\BE
\phi=\frac{1}{\sqrt{2}}\left(\phi_1+i\phi_2\right),
\EE
we find the O(2)-symmetric $\LPH$-theory Lagrangian plus coupling terms
to the gauge field:
\BE
\label{lscalar}
\ls =
\frac{1}{2} ( \partial_{\mu} \phi_1 - e_B A_{\mu} \phi_2 )^2 +
\frac{1}{2} ( \partial_{\mu} \phi_2 + e_B A_{\mu} \phi_1 )^2 -
\frac{1}{2} m_B^2(\phi_1^2 + \phi_2^2) -
\lambda_B (\phi_1^2 + \phi_2^2)^2.
\EE
Forming the Hamiltonian density
\BE
\hs \equiv \dot{\phi}_1 \Pi_1 + \dot{\phi}_2 \Pi_2 - \ls,
\EE
with $\Pi_i \equiv \delta {\cal L} / \delta \dot{\phi}_i$, we obtain:
\BE
\label{hs1}
\hs = {\cal H}_{O(2)} -
e_B \vec{A}.(\phi_1 \del \phi_2 - \phi_2 \del \phi_1) -
\frac{1}{2} e_B^2 A_{\mu} A^{\mu} ( \phi_1^2 + \phi_2^2),
\EE
where ${\cal H}_{O(2)}$ is the Hamiltonian density for
O(2)-symmetric $\LPH$ theory.

   Without loss of generality, we can choose the classical field $\PC$ to
lie in the $\phi_1$ direction.  Our trial vacuum $\VAC$
is a direct product of the free-field vacua for the $\hat{\phi}_1$ ``radial''
field (with $\hat{\phi}_1 \equiv \phi_1 - \PC$), with mass $\Omega$; for
the $\phi_2$ ``transverse'' field, with mass $\omega$; and for the gauge
fields (to be discussed below).  The middle term in (\ref{hs1})
therefore gives no contribution when we take the expectation value of
$\hs$ in the trial state $\VAC$.  Hence, we find:
\BE
\label{hs}
<\hs> \; = V_G^{O(2)} - \frac{1}{2} e_B^2 <\!A_{\mu} A^{\mu}\!>
( \PC^2 + I_{0}(\Omega) + I_0(\omega)),
\EE
where the first term is the O(2) $\LPH$-theory result \cite{brihaye,ON}:
\BA
V_G^{O(2)} & = & J(\Omega) +  J(\omega) +
\frac{1}{2} m_B^2 (\PC^2 + I_0(\Omega) + I_0(\omega))
\nonumber \\
\label{VO2}
& & \mbox{} +
\LB \left[ 3(I_0(\Omega) + \PC^2)^2 + 2 I_0(\omega) (I_0(\Omega) + \PC^2)
 + 3 I_0(\omega)^2 - 2 \PC^4 \right].
\EA

  The gauge-field Lagrangian, including gauge-fixing terms, can be
written as:
\BE
\lG = - \frac{1}{4} F_{\mu \nu} F^{\mu \nu} + (\partial_{\mu} B) A^{\mu}
+ \frac{1}{2} \xi B^2,
\EE
where $F_{\mu \nu} \equiv \partial_{\mu} A_{\nu} - \partial_{\nu} A_{\mu}$.
The last two terms, involving the
Nakanishi-Lautrup \cite{Nak} auxiliary field $B$, are equivalent to the
usual covariant gauge-fixing term $- \frac{1}{2 \xi} (\partial \cdot A)^2$,
where $\xi$ is the gauge parameter.  [To see this one integrates by parts to
get $-B (\partial \cdot A) + \frac{1}{2} \xi B^2$, and then eliminates $B$
by its equation of motion $B = (\partial \cdot A)/\xi$.]

    By itself $\lG$ would just describe a set of free massless fields.
We want to consider a generalization of this
Lagrangian that includes a mass term:
\BE
{\cal L}_{{\rm Trial}} = \lG + \frac{1}{2} \Delta^2
A_{\mu} A^{\mu}.
\EE
The ground state of this ``trial theory'' will provide us with our trial
vacuum state, with the mass $\Delta$ playing the r\^{o}le of a variational
parameter.  To construct the GEP we shall then need to take the expectation
value of $\hG$ (which we can obtain from ${\cal H}_{{\rm Trial}}$ by setting
$\Delta = 0$) in the vacuum state of ${\cal H}_{{\rm Trial}}$.

  The content of the ``trial theory'' is made plain by defining
\BE
{\cal A}_{\mu} \equiv A_{\mu} + \frac{1}{\Delta^2} \partial_{\mu}B,
\EE
which de-couples ${\cal L}_{{\rm Trial}}$ into separate ${\cal A}_{\mu}$
and $B$ sectors:
\BE
{\cal L}_{{\rm Trial}} = \left(
- \frac{1}{4} {\cal F}_{\mu\nu} {\cal F}^{\mu\nu} +
\frac{1}{2} \Delta^2 {\cal A}_{\mu} {\cal A}^{\mu} \right) -
\frac{1}{\Delta^2} \left( \frac{1}{2} \partial_{\mu}B \partial^{\mu}B -
\frac{1}{2} \xi \Delta^2 B^2 \right).
\EE
The ${\cal A}_{\mu}$ field is thus a free, massive vector field, and its
equation of motion
$\partial_{\mu} {\cal F}^{\mu\nu} + \delsq {\cal A}^{\nu} = 0$ yields
both $(\partial^2 + \Delta^2) {\cal A}^{\nu} = 0$ and
$\partial \cdot {\cal A} = 0$.
The $B$ field is a normal scalar field, mass $\sqrt{\xi} \Delta$, except
that its Lagrangian has the ``wrong sign'' and has an overall factor
$1/\Delta^2$.  It is now a relatively straightforward exercise to obtain
the Hamiltonian and canonically quantize the theory, and we just list some
of the key steps below.

  The plane-wave expansion for the ${\cal A}_{\mu}$ field is:
\BE
{\cal A}_{\mu} = \sum_{\LL} \int
\frac{{\rm d}^3 k}{(2\pi)^3 \; 2 \omega_{\underline{k}}(\Delta)}
\left[ a(\vec{k}, \LL) \epsilon_{\mu}(\vec{k}, \LL) {\rm e}^{-i k.x} +
{\rm h.c.} \right],
\EE
in which $k^0 = \omega_{\underline{k}}(\Delta) \equiv
\sqrt{\vec{k}^2 + \Delta^2}$, and the three polarization vectors
$\epsilon_{\mu}(\vec{k}, \LL)$, with $\LL = -1,0,1$ being the helicity
label, satisfy the usual completeness relation:
\BE
\sum_{\LL} \epsilon_{\mu}^{\ast}(\vec{k}, \LL) \epsilon_{\nu}(\vec{k}, \LL)
= - \left( g_{\mu\nu} - \frac{k_{\mu} k_{\nu}}{\Delta^2} \right).
\EE
The creation-annihilation operators obey
\BE
[a(\vec{k}, \LL), a^{\dagger}(\vec{k}', \LL')] =
2 \omega_{\underline{k}}(\Delta) (2 \pi)^3 \delta^{(3)}(\vec{k} - \vec{k}')
\delta_{\LL \LL'}.
\EE
The plane-wave expansion for $B$ is
\BE
B = \Delta \int
\frac{{\rm d}^3 k}{(2\pi)^3 \; 2 \omega_{\underline{k}}(\sqrt{\xi} \Delta)}
\left[ a(\vec{k}, B) {\rm e}^{-i k.x} + {\rm h.c.} \right],
\EE
in which $k^0 = \omega_{\underline{k}}(\sqrt{\xi} \Delta) \equiv
\sqrt{\vec{k}^2 + \xi \Delta^2}$, and the operators obey a ``wrong-sign''
commutation relation:
\BE
[a(\vec{k}, B), a^{\dagger}(\vec{k}', B)] = - \:
2 \omega_{\underline{k}}(\sqrt{\xi} \Delta) (2 \pi)^3
\delta^{(3)}(\vec{k} - \vec{k}').
\EE

Our trial vacuum state $\VAC$ is, by definition, annihilated by the
operators $a(\vec{k},\LL)$ ($\LL =-1,0,1$) and $a(\vec{k},B)$.
To construct the GEP we need to substitute
the above plane-wave expansions into $\hG$ (conveniently obtained from
${\cal H}_{{\rm Trial}}$ by dropping all terms involving $\Delta$) and
then sandwich the result between
$\mbox{}_{{\scriptscriptstyle G}} \!\!<\! 0\,|$ and $\VAC$.
{}From the ${\cal A}_{\mu}$ fields one obtains
\BE
<\! {\cal H}_{{\cal A}} \! > \; = 3 J(\Delta),
\EE
which is three times (because of the three polarization states) the usual
GEP result for a massless-scalar Hamiltonian evaluated in a free-field vacuum
state of mass $\Delta$.  The $B$ fields give
\BE
<\! {\cal H}_{B} \! > \; = J(\sqrt{\xi} \Delta),
\EE
which is positive because the ``wrong sign'' of the Hamiltonian is compensated
by the ``wrong sign'' of the commutator.
Therefore, in total we have
\BE
\label{hg}
<\! \hG \! > \; = 3 J(\Delta) + J(\sqrt{\xi} \Delta).
\EE

[As a check, note that if we were considering the gauge sector by itself,
minimization of (\ref{hg}) would yield $\bar{\Delta} =0$ and the result would
reduce to $4 J(0) = 4 I_1(0)$.  Recalling that the ghosts contribute
$-2 I_1(0)$, the total is $2 I_1(0)$, which is the vacuum energy associated
with two massless, bosonic degrees of freedom.  These correspond to the two
transverse polarizations of the massless vector field.  Thus, we see
explicitly, for any value of the gauge parameter $\xi$, how the ghosts act
to cancel out the vacuum-energy contributions from the unphysical components
of the gauge field \cite{ambj}.]

   To obtain the total GEP we combine $<\! \hG \!>$ from (\ref{hg}) with
$<\! \hs \!>$ from (\ref{hs}).  A short calculation gives:
\BA
<\! A_{\mu}A^{\mu} \!> &=& < \! {\cal A}_{\mu} {\cal A}^{\mu} \!> +
\frac{1}{\Delta^4} < \! \partial_{\mu} B \partial^{\mu} B \! >
\nonumber \\
&=& -3 I_0(\Delta) - \xi I_0(\sqrt{\xi} \Delta),
\EA
so we obtain finally:
\BA
V_G(\PC;\Omega,\omega,\Delta) &=&
V_G^{O(2)} + 3 J(\Delta) + J(\sqrt{\xi} \Delta)
\nonumber \\
\label{result}
& & \mbox{} + \frac{1}{2} e_B^2 (3 I_0(\Delta) + \xi I_0(\sqrt{\xi} \Delta) )
(\PC^2 + I_0(\Omega) + I_0(\omega) ).
\EA

Minimization with respect to the mass parameters $\Delta$, $\Omega$, and
$\omega$ leads to:
\BE
\label{deq}
{\bar{\Delta}}^2= e_B^2 [ \varphi^2_c + I_0(\bar \Omega) + I_0(\bar \omega)],
\EE
\BE
\label{Oeq}
{\bar \Omega}^2= m_B^2 + 4 \lambda_B [ 3 I_0(\bar \Omega) + I_0(\bar \omega)
+ 3 \varphi^2_c] +
e^2_B [ 3 I_0(\bar \Delta) + \xi I_0(\sqrt{\xi} \bar\Delta) ],
\EE
\BE
\label{oeq}
{\bar \omega}^2= m_B^2 + 4 \lambda_B [  I_0(\bar \Omega) + 3 I_0(\bar \omega)
+ \varphi^2_c] +
e^2_B [ 3 I_0(\bar \Delta) + \xi I_0(\sqrt{\xi} \bar\Delta) ].
\EE
%
%
%
%
\section{Covariant $\delta$-expansion Calculation}
\setcounter{equation}{0}

In this section we perform the calculation in the Euclidean
functional-integral formalism in the manner of Ref. \cite{paper1}.  Note
that in passing to the Euclidean formalism the Minkowski scalar product
$a^{\mu}b_{\mu}$ goes to $-a_{\mu}b_{\mu}$; thus terms with just one pair of
contracted indices change sign relative to other terms.  The Euclidean action
reads:
\BA
S & = & \int d^{4}x \;\left[\frac{1}{4}\:F_{\mu\nu}F_{\mu\nu} +
\frac{1}{2\xi}\left(\partial_{\mu}A_{\mu}\right)^2
\right.
\nonumber
\\
& & \hspace{15mm}\left.\frac{}{} +
\left(D_{\mu}\phi\right)^{\ast}\left(D_{\mu}\phi\right)
+ m_B^2\phi^{\ast}\phi + 4\lambda_B \left(\phi^{\ast}\phi\right)^2\right].
\EA
Rewritten in terms of the real scalar fields $\phi_1$ and $\phi_2$, the
action becomes
\BA
S & = &
\int d^{4}x \left[ \:\frac{1}{4}\:F_{\mu\nu}F_{\mu\nu} +
\frac{1}{2\xi}\left(\partial_{\mu}A_{\mu}\right)^2
+\frac{1}{2}\:\phi_1\left(-\partial^2+m_B^2\right)\phi_1
\right.
\nonumber
\\
& & \hspace{12mm}\mbox{}
+\frac{1}{2}\:\phi_2\left(-\partial^2+m_B^2\right)\phi_2
+ \lambda_B \left(\phi^{2}_{1}+\phi^{2}_{2}\right)^2
\nonumber
\\
& & \hspace{12mm}\mbox{}\left.
+ \frac{1}{2}\:e_B^2 A_{\mu}A_{\mu}\left(\phi^{2}_{1}+\phi^{2}_{2}\right) +
e_B A_{\mu}\left(\phi_1\partial_{\mu}\phi_2-\phi_2\partial_{\mu}\phi_1\right)
\right],
\EA
where $\partial^2 \equiv \partial_{\mu}\partial_{\mu}$.
Introducing a source for the $\phi_1$ field, the generating functional
is given by:
\BE
\label{zj}
Z[j]=\int D\left[\phi_1,\phi_2,A_{\mu} \right] \;
\exp \left(\mbox{}-S + \int \! d^4 x \;j(x)\phi_1(x) \right),
\EE
and the effective action is obtained by the Legendre transformation
\BE
\label{lt}
\Gamma[\PC]=\ln Z[j] - \int d^{4}x\;j(x)\PC(x),
\EE
where the classical field, $\PC(x)$,
is the vacuum expectation value of the field $\phi_1(x)$ in the presence
of the source $j(x)$. The effective potential itself, $V_{{\rm eff}}(\PC)$,
is obtained from $\Gamma[\PC]$  by setting $\PC(x)$ to be a constant,
$\PC$, and dividing out a minus sign and a spacetime volume factor.

Generalizing the procedure of Ref. \cite{paper1} (see also
\cite{oko,delta,gandhi}), we can calculate the GEP from a first-order
expansion in a nonstandard kind of perturbation theory.  First we introduce
the shifted fields:
\BE
\hat{\phi}_{1}(x) = \phi_{1}(x) - \PC,\hspace {8mm}
\hat{\phi}_{2}(x) = \phi_{2}(x).
\EE
(Notice that we have taken the shift parameter to be exactly $\PC$,
the vacuum expectation value of $\phi(x)$:  Although not obligatory
\cite{paper1}, this simplifies the calculation.)

We then split the (Euclidean) Lagrangian into two parts:
\BE
{\cal L} = \left( \frac{}{}{\cal L}_{0} + {\cal L}_{int}\right)_{\delta=1},
\EE
where ${\cal L}_{0}$ is a sum of three free-field Lagrangians: one for
the vector field $A_{\mu}$, of mass $\Delta$; one for the radial scalar field,
$\hat{\phi}_{1}$, with mass $\Omega$; and one for the transverse scalar
field, $\hat{\phi}_{2}$, with mass $\omega$:
\BA
{\cal L}_{0} & = & \frac{1}{2} \; A_{\mu}(x)
\left[\left(-\partial^2+\Delta^2\right)
\delta_{\mu \nu}+\left(1-\frac{1}{\xi}\right)\partial_{\mu}\partial_{\nu}
\right] A_{\nu}(x)
\nonumber
\\
& & \mbox{}
+ \frac{1}{2} \; \hat{\phi}_{1}(x)
\left(-\partial^2+\Omega^{2}\right)\hat{\phi}_{1}(x)
+ \frac{1}{2} \; \hat{\phi}_{2}(x)
\left(-\partial^2+\omega^{2}\right)\hat{\phi}_{2}(x).
\EA
The interaction Lagrangian is then:
\BA
{\cal L}_{int} & = &
\delta \left[ \frac{}{} v_{0} + v_{1}\hat{\phi}_{1} + v_{2}\hat{\phi}_{1}^{2}
+ v_{3}\hat{\phi}_{1}^{3} + \lambda_{B}\hat{\phi}_{1}^{4}
+ v'_{2}\hat{\phi}_{2}^2 + \lambda_{B}\hat{\phi}_{2}^4
\right.
\nonumber
\\
& & \mbox{}
\hspace{5mm}
+ 4 \lambda_{B} \PC \; \hat{\phi}_{1} \hat{\phi}_{2}^2
+ 2 \lambda_{B} \; \hat{\phi}_{1}^2 \hat{\phi}_{2}^2
+ \frac{1}{2}\left(e_B^2\PC^2-\Delta^2\right)A_{\mu}A_{\mu}
\nonumber
\\
& & \mbox{}
\hspace{5mm}
+ e_{B}\PC\:A_{\mu}\:\partial_{\mu}\hat{\phi}_{2}
+ e_{B}A_{\mu}\left(\hat{\phi}_{1}\partial_{\mu}\hat{\phi}_{2}
                    -\hat{\phi}_{2}\partial_{\mu}\hat{\phi}_{1}\right)
\nonumber
\\
& & \mbox{}
\hspace{5mm}
+ e_{B}^2\PC\:\hat{\phi}_{1}A_{\mu}A_{\mu} + \frac{1}{2}e_{B}^2
A_{\mu}A_{\mu}\left(\hat{\phi}_{1}^2+\hat{\phi}_{2}^2\right)
\left. \frac{}{} \!\!\right].
\label{eq:deflint}
\EA
The ``coupling constants'' $v_{0},v_{1},v_{2},v'_{2}$ and $v_{3}$, which are
$\PC$ dependent, are the same as in the $\LPH$ case \cite{paper1}:
\BA
v_{0} & = & \frac{1}{2}m_{B}^{2}\PC^{2}+\lambda_{B}\PC^{4},
\nonumber \\
v_{1} & = & \left( m_{B}^{2}+4\lambda_{B}\PC^{2}\right)\PC,
\nonumber \\
v_{2} & = & \frac{1}{2}\left(m_{B}^{2}-\Omega^{2}\right)
            +6\lambda_{B}\PC^{2},
\\
v'_{2} & = & \frac{1}{2}\left(m_{B}^{2}-\omega^{2}\right)
             +2\lambda_{B}\PC^{2},
\nonumber \\
v_{3} & = & 4\lambda_{B}\PC .
\nonumber
\EA

The artificial expansion parameter $\delta$ has been introduced in
${\cal L}_{int}$ in order to keep track of the order of approximation, which
consists in obtaining a (truncated) Taylor series in $\delta$, about
$\delta=0$, which is then used to extrapolate to $\delta=1$.

The expansion in ${\cal L}_{int}$ (or equivalently in $\delta$) is now quite
straightforward, following standard perturbation theory procedures.  To first
order in $\delta$ it yields:
\BA
\Gamma[\PC] & = & \Gamma^{O(2)}[\PC]
-\frac{1}{2}\:\mbox{Tr}\:\ln\left[G_{\mu\nu}^{-1}(x,y)\right]
\nonumber
\\
& &\mbox{}-\delta\int \!{\rm d}^4 x \:
\left\{\frac{1}{2}\left(e_B^2\PC^2-\Delta^2
\right)G_{\mu\mu}(x,x)
\right.
\nonumber
\\
& &\left.\hspace{31mm}\mbox{}
+\frac{1}{2}e_B^2\left[I_0(\Omega)+I_0(\omega)\right]G_{\mu\mu}(x,x)\right\}
+O\left(\delta^2\right),
\EA
where $\Gamma^{O(2)}[\PC]$ is the first-order action for the scalar sector,
and
\BE
G_{\mu\nu}(x,y)  =  \int \! \frac{{\rm d}^4 p}{(2 \pi)^4}
\:\frac{1}{p^2+\Delta^2}
\left[\delta_{\mu\nu}+(\xi-1)\frac{p_{\mu}p_{\nu}}{p^2+\xi\Delta^2}\right]
\:e^{-ip\cdot(x-y)}
\EE
is the gauge field propagator, $<\!A_{\mu}(x)A_{\nu}(y)\!>$.
The Trace-log term gives
\BE
\label{trln}
\mbox{Tr}\:\ln\left[G_{\mu\nu}^{-1}(x,y)\right] =
2{\cal V}\:\left[3I_1(\Delta)+I_1(\sqrt{\xi}\:\Delta)\right] ,
\EE
where ${\cal V}$ is the infinite spacetime volume, and the contracted
propagator gives
\BE
G_{\mu\mu}(x,x) =  3I_0(\Delta)+\xi I_0(\sqrt{\xi}\:\Delta).
\EE

To obtain the GEP we discard the $O(\delta^2)$ terms and set $\delta = 1$
and divide through $-{\cal V}$ to obtain
\BA
V_G = V_G^{O(2)} & & \mbox {\hspace*{-6mm}} +
3I_1(\Delta) + I_1(\sqrt{\xi}\Delta) +
\frac{1}{2}\left(e_B^2\PC^2-\Delta^2 \right)
\left( 3I_0 (\Delta)+\xi I_0(\sqrt{\xi}\:\Delta) \right)
\nonumber \\
& & \mbox {\hspace*{-6mm}}+
\frac{1}{2}e_B^2 \left[ I_0(\Omega)+I_0(\omega) \right]
\left( 3I_0(\Delta)+\xi I_0(\sqrt{\xi}\:\Delta) \right).
\EA
Recalling that $J(\Delta)\equiv I_1(\Delta)-\frac{1}{2}\Delta^2 I_0(\Delta)$,
one sees that this result coincides with the result obtained from the
canonical calculation, Eq. (\ref{result}).
%
%
%
%
\section{Covariant Variational Calculation}
\setcounter{equation}{0}

  In this section we use the method developed in \cite{jensen} based on
Feynman's variational principle \cite{feyn} applied to the Euclidean action.
This in turn follows from Jensen's inequality for expectation values of
convex functions; in particular, exponential functions:
\BE
\int {\rm d} \mu(\phi) \exp g(\phi) \geq
\exp \left( \int {\rm d} \mu(\phi) g(\phi) \right),
\EE
for a normalized integration measure ${\rm d} \mu(\phi)$.
The inequality applies only to commuting fields, but happily in the U(1)
case the anticommuting ghost fields can be integrated out exactly.
The remaining action can be written in the following form (using the shifted
fields $\PHA = \phi_{1} - \PC$, $\PHB = \phi_{2}$, as in the last section):
\BE
	S[A_{\mu},\PHA,\PHB]= S_A[A_\mu]
+ S_{A,\phi} [A_\mu,\PHA,\PHB] + S_{\phi}[\PHA,\PHB]\>,
\EE
where
\BE
	S_A = \frac{1}{2} \int \! {\rm d}^4 x
 A_{\mu}(x) \left [
- \partial^2 \delta_{\mu \nu}
+ (1- \frac{1}{\xi}) \partial_\mu \partial_\nu \right] A_\nu(x) ,
\EE
\BA
S_{A,\phi} &=& \int \! {\rm d}^4 x
\left\{ \frac{1}{2} e_B^2 (\varphi_c^2 + \PHA^2 + \PHB^2)
A_{\mu}(x) A_{\mu}(x) \right. \nonumber
\\
& & + \left. e_B A_{\mu} \left[
(\PHA+\PC) \partial_{\mu} \PHB +
\PHB \partial_{\mu} (\PHA+\PC) \right ]
+ e_B^2 \PC \PHA A_{\mu} A_{\mu} \right \},
\EA
and $S_{\phi}$ is given by the usual O(2) $\LPH$ action.

   Following Ref. \cite{jensen}, we now apply the Feynman-Jensen inequality
to $Z[j]$, Eq (\ref{zj}), with
${\rm d} \mu(\phi) = {\cal N}^{-1} D A_{\mu} D\PHA D\PHB {\rm e}^{-S_G}$ and
$g(\phi) = S_G - S + \int j \phi$, where
\BE
{\cal N}= \int D A_{\mu} D\PHA D\PHB \>{\rm e}^{-S_G},
\EE
and where $S_G$ is a quadratic ``trial action'':
\BE
S_G = \frac{1}{2} \int {\rm d}^4 x \left\{
A_{\mu} G^{-1}_{\mu \nu} A_{\nu}+
\PHA G_1^{-1} \PHA + \PHB G_2^{-1} \PHB \right\},
\EE
involving adjustable kernels $G^{-1}$.  Taking the Legendre transform,
(\ref{lt}), we obtain the ``Gaussian effective action'' \cite{jensen}:
\BE\label{aefa}
\bar{\Gamma}^{GEA}[\varphi_c]
 = \max_{G} \left \{
\log({\cal N})+ {\cal N}^{-1}\int\>D A_{\mu} D\PHA D\PHB
\>{\rm e}^{-S_G}\>(S_G-S)\right\},
\EE
as a lower bound on the exact effective action (which will hence yield an
upper bound on the effective potential).  Since the kernels involve
differential operators, it is convenient to go to momentum space, using
Fourier transforms (indicated by tildes) and the convenient notation
$\int_p = \int \,{\rm d}^4 p/(2\pi)^4$, $\>\bar\delta(p)=(2\pi)^4\delta(p)$.
We can then write the trial action as
\BE
S_G = \frac{1}{2} \int_p \int_q \left\{
{\tilde A}_{\mu}(p) {\tilde G}^{-1}_{\mu \nu}(p,q) {\tilde A}_{\nu}(q) +
\tilde{\phi}_1(p) {\tilde{G}}_1^{-1}(p,q) \tilde{\phi}_1(q)
 + \tilde{\phi}_2(p) {\tilde{G}}_2^{-1}(p,q) \tilde{\phi}_2(q) \right\},
\EE
in which the $\tilde{G}^{-1}$'s are the inverses of the momentum-space
propagators.

   Evaluation of the Gaussian functional integrals involved in (\ref{aefa})
is straightforward, and yields
\[
\bar{\Gamma}^{GEA} =  \max_{{\tilde G}} \left\{
\Gamma^{O(2)}[\TF,{\tilde G}_1,{\tilde G}_2]
 - \frac{1}{2} {\rm Tr}\> \ln \, [{\tilde G}^{-1}_{\mu\nu}(p,q)]
\right. \]
\[ - \frac{1}{2} \int_{p}
 \left [ p^2 \delta_{\mu \nu} +
e^2_B
\int_r({\tilde G}_1(r,-r) + {\tilde G}_2(r,-r))\delta_{\mu \nu}
-(1 - \frac{1}{\xi}) p_{\mu} p_{\nu}\right ]
{\tilde G}_{\mu \nu}(p,-p) \]
\BE
\label{geaeq}
- \left. \frac{1}{2} e^2_B
\int_{pqrs}\TF(r) \TF(s) {\tilde G}_{\mu \mu}(p,q)
\bar\delta(p+q+r+s) \right\} .
\EE
Maximization yields optimization equations determining the optimal
${\tilde G}$ propagators, denoted by ${\bar G}(p,q)$:
\BA
\label{deq1}
{\bar G}^{-1}_{\mu\nu}(p,q) &=&
 \left [ p^2 \delta_{\mu \nu} + e^2_B
\int_r({\bar G}_1(r,-r) + {\bar G}_2(r,-r))\delta_{\mu \nu}
-(1 - \frac{1}{\xi}) p_{\mu} p_{\nu}\right ] \bar\delta(p+q)
\nonumber \\
& & \mbox{} + e^2_B \int_{rs}\TF(r) \TF(s) \bar\delta(p+q+r+s)
\delta_{\mu \nu},
\\
{\bar G}^{-1}_1(p,q) &=&
\left [ p^2 + m_B^2 + e^2_B
\int_r {\bar G}_{\mu\mu} (r,-r) \right ] \bar\delta(p+q)
\nonumber \\
& & \mbox{} + 4 \lambda_B \int_{rs}
[3 {\bar G}_1(r,s) + {\bar G}_2(r,s)  + 3 \TF(r) \TF(s)]
\bar\delta(p+q+r+s),
\\
{\bar G}^{-1}_2(p,q) &=&  \left [ p^2 + m_B^2
 +e^2_B \int_r {\bar G}_{\mu\mu} (r,-r)  \right ] \bar\delta(p+q)
\nonumber \\
& & \mbox{} + 4 \lambda_B \int_{rs}
[ {\bar G}_1(r,s) + 3 {\bar G}_2(r,s)  +  \TF(r) \TF(s)]
\bar\delta(p+q+r+s) .
\EA
For a spatially constant classical field we have
$\TF(p)= \varphi_c \bar\delta(p)$, and the above equations then dictate that
the propagators all become proportional to $\bar\delta(p+q)$, so we may
write them in the form
\BE
{\bar G}^{-1}_{\mu\nu}(p,q) = \left [ (p^2 + {\bar \Delta}^2)\delta_{\mu \nu}
- (1- \frac{1}{\xi}) p_{\mu}p_{\nu} \right] \bar\delta(p+q),
\EE
\BE
{\bar G}_1^{-1}(p,q) = (p^2 +{ \bar \Omega}^2)\bar\delta(p+q),
\EE
\BE
{\bar G}_2^{-1}(p,q) = (p^2 + {\bar \omega}^2)\bar\delta(p+q),
\EE
where the optimal mass parameters $\bar{\Delta}$, $\bar \Omega$, and
$\bar \omega$ are given by
\BE
{\bar{\Delta}}^2= e_B^2 [ \varphi^2_c + I_0(\bar \Omega) + I_0(\bar \omega)],
\EE
\BE
{\bar \Omega}^2= m_B^2 + 4 \lambda_B [ 3 I_0(\bar \Omega) + I_0(\bar \omega)
+ 3 \varphi^2_c] +
e^2_B{\cal V}^{-1}\int_p {\bar G}_{\mu\mu}(p,-p),
\EE
\BE
{\bar \omega}^2= m_B^2 + 4 \lambda_B [  I_0(\bar \Omega)
+ 3 I_0(\bar \omega) + \varphi^2_c]
          +
e^2_B{\cal V}^{-1}\int_p {\bar G}_{\mu\mu}(p,-p).
\EE
As usual, factors of ``$\bar \delta(0)$'' have been interpreted as spacetime
volume factors ${\cal V}$.  The integral $\int_p {\bar G}_{\mu\mu}(p,-p)$ ,
where
\BE
{\bar G}_{\mu\nu}(p,-p)
= \frac{{\cal V}}{p^2+ {\bar \Delta}^2}
\left [ \delta_{\mu \nu} + (\xi -1) \frac{p_\mu p_\nu}{p^2
+ \xi {\bar \Delta}^2}
\right ],
\EE
can be evaluated in terms of $I_0$ integrals:
\BE
\int_p {\bar G}_{\mu\mu}(p,-p)= {\cal V}
[ 3I_0(\bar\Delta) + \xi I_0(\sqrt{\xi} \bar\Delta)].
\EE
The Trace-log term can be taken from Eq. (\ref{trln}), so that we obtain
finally the same result as in Eq. (\ref{result}).

%
%
\section{Comments on the Unrenormalized Result}
\setcounter{equation}{0}

  The Gaussian-approximation result shows a dependence upon the gauge
parameter $\xi$.  This means that our Gaussian approximation does not fully
respect gauge invariance.  However, we argue that this is inevitable and not
fatal.  It is inevitable because, for the O(2) scalars, we have to use
``Cartesian-coordinate'' fields $\hat \phi_1$, $\hat \phi_2$ rather than
``polar-coordinate'' fields, so that, when $\PC \neq 0$, the O(2) symmetry
is not being fully respected.  In pure $\LPH$ theory this produces an
apparent conflict with Goldstone's theorem, in that the transverse mass
parameter, $\OL$, is non-zero \cite{ON,brihaye}.  However, the point is that
the transverse field $\hat \phi_2$ is not the true ``polar-angle'',
Goldstone field.  In the U(1) Higgs model, in a covariant gauge, the Goldstone
field becomes an unphysical degree of freedom \cite{higgs}, but the problem
remains in the form of gauge-parameter dependence.  This just means, though,
that we have a ``non-invariant'' approximation -- which is where the exact
result is known to be independent of some parameter, but the approximate
result has a dependence on that parameter.  This is actually quite a common
occurrence, and can be dealt with by ``optimizing'' the unphysical parameter;
requiring the approximate result to be stationary, or more generally
``minimally sensitive,'' to the unphysical parameter \cite{OPT}.  One is in
a still better position when the approximation has a variational character,
because then the optimal choice for the unphysical parameter is unquestionably
determined by minimization.

  Our calculation here does indeed have a variational character.  [One might
well have been unsure, with the canonical calculation alone, whether or
not the variational inequality is valid in the presence of a
``wrong-sign'' field (and hence negative-norm states).  However, this doubt
is allayed by the covariant variational calculation: the Jensen inequality
just depends on the the convexity of the exponential function and is equally
valid for $\exp(-g(\phi))$ and $\exp(+g(\phi))$.]  By differentiating
$\bar V_G$ one finds that the optimal gauge is the Landau gauge, $\xi = 0$.
This can easily be seen by noting that, by virtue of the $\bar{\Delta}$
equation, the $\xi$-dependence in $\bar{V}_G$ comes only from an
$I_1(\sqrt{\xi} \bar{\Delta})$ term.  Since $I_1$ is (formally) an increasing
function of its argument, the energy is minimized when $\xi = 0$.

  With $\xi=0$, and discarding a vacuum-energy contribution $I_1(0)$, the
GEP and its optimization equations simplify to:
\BE
\label{result0}
V_G(\PC;\Omega,\omega,\Delta) =
V_G^{O(2)} + 3 J(\Delta) +
\frac{3}{2} e_B^2 I_0(\Delta) (\PC^2 + I_0(\Omega) + I_0(\omega) ),
\EE
with $V_G^{O(2)}$ given by (\ref{VO2}), and
\BE
\label{deq0}
{\bar{\Delta}}^2= e_B^2 [ \varphi^2_c + I_0(\bar \Omega) + I_0(\bar \omega)],
\EE
\BE
\label{Oeq0}
{\bar \Omega}^2= m_B^2 + 4 \lambda_B [ 3 I_0(\bar \Omega) + I_0(\bar \omega)
+ 3 \varphi^2_c] +
3 e^2_B I_0(\bar \Delta),
\EE
\BE
\label{oeq0}
{\bar \omega}^2= m_B^2 + 4 \lambda_B [  I_0(\bar \Omega) + 3 I_0(\bar \omega)
+ \varphi^2_c] +
3 e^2_B I_0(\bar \Delta).
\EE
Note that if the $\bar{\Delta}$ equation, (\ref{deq0}), is substituted back
into $V_G$, (\ref{result0}), then we can write the GEP as
\BE
\label{vbar}
\bar{V}_G(\PC) = V_G^{O(2)} + 3 I_1(\bar{\Delta}),
\EE
with separate contributions from the scalar and gauge sectors.  This
observation applies to the Gaussian effective action, too, since
Eq. (\ref{deq1}) substituted back into (\ref{geaeq}) yields
\BE
\bar{\Gamma}^{GEA} =
\Gamma^{O(2)} - \frac{1}{2} {\rm Tr}\> \ln [{\bar G}^{-1}_{\mu\nu}(p,q)]\>.
\EE
Note, however, that the optimization equations for $\bar{\Delta}$,
$\bar{\Omega}$, and $\bar{\omega}$, (\ref{deq0}--\ref{oeq0}), remain
coupled.

   We may also remark that the generalization of the result to $\nu+1$
dimensions is trivial: the integrals need to be re-defined in an obvious
way, and the factors of 3 associated with the $\Delta$ integrals need to be
replaced by $\nu$, since these factors correspond to the number of
polarization states of a massive vector field.

   Finally, we briefly comment upon some previous work relating to the GEP
and the U(1) Higgs model.  (i) All\`{e}s and Tarrach \cite{at} used a
somewhat na{i}ve canonical approach which we believe is valid in Feynman
gauge ($\xi \! = \! 1$) only.  Their treatment of the scalar sector
effectively sets $\OL \equiv \OC$, which is sub-optimal.  In renormalizing
their result, All\`{e}s and Tarrach used a generalization of the
``precarious'' $\LPH$ theory, which does not have SSB.  (ii)  Cea \cite{cea}
describes a temporal-gauge GEP calculation, but contents himself with
demonstrating that the 1-loop terms are recovered correctly.  (iii)  The
papers of Ref. \cite{mt} make a comprehensive study of the Schr\"{o}dinger
wavefunctional formalism, and try hard to maintain gauge invariance and
compliance with the Goldstone theorem.  Our view, as discussed above, is
less puritanical.  (iv) Kovner and Rosenstein \cite{kr} use yet another
formulation of the Gaussian approximation, based on truncating the
Dyson-Schwinger equations.  Their renormalization is quite different from
ours:  it appears to be related to the ``precarious'' $\LPH$ renormalization,
but it somehow transfers the negative sign from $\LB$ to wavefunction
renormalization factors.

  The ``precarious'' renormalization of the U(1)-Higgs-model GEP is a
topic which we do not pursue here, but it could be of theoretical
interest:  We would expect the results to be similar to
the $1/N$-expansion analysis of Kang \cite{kang}.
%
%

\section{``Autonomous'' Renormalization of the GEP}
\setcounter{equation}{0}

\subsection{The divergent integrals}

The GEP involves the quartically and quadratically divergent integrals
$I_1$ and $I_0$.  Another related integral:
\BE
I_{-1}(\OC) \equiv -2 \frac{{\rm d} I_0}{{\rm d} \OC^2}
= \int \frac{{\rm d}^3 p}{(2 \pi)^3}
\frac{1}{2 (\omega_{\underline{p}})^3}
= 2 \int \frac{{\rm d}^4 p}{(2 \pi)^4}
\frac{1}{(p^2 + \Omega^2)^2},
\EE
which is logarithmically divergent, will play a crucial role.  Ref. \cite{gep2}
derives useful formulas for these integrals by Taylor-expanding the integrands
about $\OC^2 = m^2$, and then re-summing the terms that give convergent
integrals.  From these we can obtain the still more convenient formulas:
\BE
\label{i1key}
I_1(\OC) = I_1(0) + \frac{\OC^2}{2} I_0(0) - \frac{\OC^4}{8} I_{-1}(\mu)
+ f(\OC^2),
\EE
\BE
\label{key}
I_0(\OC) = I_0(0) - \frac{\OC^2}{2} I_{-1}(\mu) + 2 f'(\OC^2),
\EE
\BE
\label{key2}
I_{-1}(\OC) = I_{-1}(\mu) - \frac{1}{8 \pi^2} \ln \frac{\OC^2}{\mu^2},
\EE
where
\BE
\label{f}
f(\OC^2) = \frac{\OC^4}{64 \pi^2} \left [\ln \frac {\OC^2}{\mu^2} -
\frac{3}{2}\right],
\EE
and $f'(\OC^2)$ is its derivative with respect to $\OC^2$.  These formulas
are valid in any regularization scheme that preserves the property
${\rm d}I_n/{\rm d} \OC^2 = (n-\half) I_{n-1}$.  This allows one, at
least in the $\LPH$ case, to discuss the renormalization procedure in a
completely regularization-independent manner.  However, in gauge theories,
most cutoff-based renormalizations -- because they interfere with gauge
invariance -- have problems with quadratic divergences in the vector
self-energy.  Here these problems would manifest themselves as quadratic
divergences in the vector-mass parameter $\bar \Delta^2$, Eq. (\ref{deq0})
\cite{mt}.  (Unlike the scalar case, these cannot be simply absorbed into a
bare-mass.)  It is well known in other contexts that, with sufficient technical
virtuosity, these problems can be shown to be spurious \cite{schwinger}.
However, it is much simpler to appeal to dimensional regularization, or some
such scheme, in which one can justify setting the scale-less integrals,
$I_0(0)$ and $I_1(0)$, equal to zero.  This automatically elimates any problem
with quadratic divergences.  All the remaining divergences can be written in
terms of $I_{-1}$, which has a $1/\epsilon$ pole in dimensional regularization:
Explicitly:
\BE
I_{-1}(\OC) = \frac{A}{\epsilon}\OC^{-\epsilon}, \quad\quad
A \equiv \frac{1}{4 \pi^2} \Gamma(1+\epsilon/2)(4 \pi)^{\epsilon/2}.
\EE

\subsection{Renormalization: Part I}

To renormalize $\bar V_G$ we use an ``autonomous'' renormalization
(Cf. \cite{auto2,ON}), characterized by an infinite re-scaling of the classical
field and infinitesimal bare coupling constants:
\BE
\label{autren1}
\varphi^2_c = Z_\phi \Phi_c^2 = z_0 I_{-1}(\mu) \Phi_c^2,
\nonumber
\EE
\BE
\label{autren2}
\lambda_B = \frac{\eta}{I_{-1}(\mu)}\>,
\qquad
e^2_B= \frac{\gamma}{I_{-1}(\mu)}\>,
\EE
where $z_0$, $\eta$ and $\gamma$ are finite, and $\mu$ is a finite mass scale.
For the present, we assume that all the $I_{-1}$ factors have the same
argument, $\mu$.  We shall also take $m_B^2 =0$.  These simplifying
assumptions will be removed later in subsection 7.4.  We shall also
postpone the determination of the finite wavefunction-renormalization factor
$z_0$ to that subsection.

  First, we substitute the renormalization equations into the
optimization equations (\ref{deq0} -- \ref{oeq0}) and use the key formula
for $I_0$, (\ref{key}), setting $I_0(0)=0$.  Keeping only the finite terms,
for the present, we obtain:
\BA
\bar\Delta^2 &=& \gamma ( z_0\Phi_c^2 - \half \bar\OC^2 - \half \bar\OL^2 )
+ \epsilon_{\Delta},
\nonumber \\
\label{gap1}
\bar\OC^2 &=& 4\eta (- \frac{3}{2} \bar\OC^2 - \half \bar\OL^2 +
3 z_0 \Phi_c^2 ) - \frac{3}{2} \gamma \bar\Delta^2
+ \epsilon_{\OC},
\\
\bar\OL^2 &=& 4\eta (- \half \bar\OC^2 - \frac{3}{2} \bar\OL^2 + z_0 \Phi_c^2 )
- \frac{3}{2} \gamma \bar\Delta^2
+ \epsilon_{\OL},
\nonumber
\EA
where the $\epsilon_{\Delta}$, $\epsilon_{\OC}$, $\epsilon_{\OL}$ terms
are infinitesimal, ${\cal O}(1/I_{-1})$, terms.  Ignoring the $\epsilon$
terms the equations are linear and homogeneous, so that each mass parameter
is proportional to $\Phi_c$.  The equations can be straightforwardly
solved to yield:
\BA
\bar\Delta^2 &=& \frac{2 \gamma}{(2+16\eta-3\gamma^2)} z_0 \Phi_c^2
 + {\cal O}(1/I_{-1}),
\nonumber \\
\label{gap2}
\bar\OC^2 &=& \frac{[8\eta(3+16\eta)-3\gamma^2(1+8\eta)]}
{(1+4\eta)(2+16\eta-3\gamma^2)} z_0 \Phi_c^2
 + {\cal O}(1/I_{-1}),
\\
\bar\OL^2 &=& \frac{(8\eta-3\gamma^2)}{(1+4\eta)(2+16\eta-3\gamma^2)}
z_0 \Phi_c^2
 + {\cal O}(1/I_{-1}).
\nonumber
\EA

Since $\partial V_G/\partial \OC =0$, etc., by virtue of the gap
equations, the total derivative of $\bar{V}_G$ with respect to $\PC$
is equal to its partial derivative, and so can be calculated very easily:
\BA
\frac{{\rm d} \bar V_G} {{\rm d} \PC} =
\frac{\partial V_G} {\partial \PC}
& = &  \PC [ m_B^2 + 4 \LB(3I_0(\bar\OC)+I_0(\bar\OL)+\PC^2) +
3 e_B^2 I_0(\bar\Delta) ]
\nonumber \\
\label{dvdphi}
& = & \PC (\bar\OC^2 - 8 \LB  \PC^2).
\EA
The last equality follows from the optimization equation for $\bar\OC$,
(\ref{Oeq0}), and yields the same expression as in pure $\LPH$ theory
\cite{ON}.  In order for $\bar{V}_G$ to be finite in terms of the
re-scaled field $\Phi_c$, we must have a cancellation between the finite
part of $\bar\OC^2$ and $8 \LB  \PC^2 = 8 \eta z_0 \Phi_c^2$.  This condition
implies a constraint on the coefficients $\eta$ and $\gamma$ of the $\LB$ and
$e_B^2$ coupling constants.  This can be expressed as:
\BE
\label {re2}
\gamma^2 = \frac{8 \eta}{3} \frac{(1 - 8 \eta - 64 \eta^2)}{(1-32 \eta^2)},
\EE
and will be discussed further in the next subsection.

Using the constraint one can simplify the expressions for the optimal mass
parameters to:
\BA
\bar \Delta^2  &=& \gamma \frac{1- 32 \eta^2}{1+ 4 \eta} z_0 \Phi_c^2
 + {\cal O}(1/I_{-1}),
\nonumber \\
\label {re1}
\bar \Omega^2  &=& 8 \eta z_0 \Phi_c^2
 + {\cal O}(1/I_{-1}),
\\
\bar \omega^2  &=& \frac{ 32 \eta^2}{1+ 4 \eta} z_0 \Phi_c^2
 + {\cal O}(1/I_{-1}).
\nonumber
\EA

   The renormalized GEP is most easily obtained from the expression
for its first derivative, (\ref{dvdphi}).  The leading terms cancel, so
one needs to obtain the infinitesimal, ${\cal O}(1/I_{-1})$, part of
$\bar \OC^2$.  The calculation is straightforward, if tedious.  One needs
to obtain the explicit form of $\epsilon_{\Delta}, \epsilon_{\OC},
\epsilon_{\OL},$ in Eqs. (\ref{gap1}) by going back to Eqs.
(\ref{deq0} -- \ref{oeq0}).  One can then solve for the ${\cal O}(1/I_{-1})$
correction to $\bar \OC$ in Eq. (\ref{re1}).  After some algebra, one
finds that the coefficients of the three $f'$ terms match those in (\ref{re1})
above, so that one can write:
\BE
\frac{{\rm d} \bar V_G} {{\rm d} \Phi_c} = 2 \Phi_c \left[
3 \left( \frac{{\rm d} \bar \Delta^2} {{\rm d} \Phi_c^2} \right)
f'(\bar \Delta^2)
+ \left( \frac{{\rm d} \bar \OC^2}    {{\rm d} \Phi_c^2} \right)
f'(\bar \OC^2)
+ \left( \frac{{\rm d} \bar \OL^2}    {{\rm d} \Phi_c^2} \right)
f'(\bar \OL^2)
\right] .
\EE
Thus, by integrating with respect to $\Phi_c$, one obtains the renormalized
GEP as just:
\BE
\label{rgep}
\bar V_{G}= 3 f(\bar \Delta^2) + f(\bar \Omega^2) + f(\bar \omega^2) ,
\EE
where $f$ is the function defined in Eq. (\ref{f}).  The GEP is thus
a sum of $\Phi_c^4 \ln \Phi_c^2$ and $\Phi_c^4$ terms.  If we swap the
parameter $\mu$ for the vacuum value $\Phi_v$ (defined as the position of
the minimum of $\bar V_{G}$), we can write the GEP simply as
\BE
\label{vgv}
\bar V_{G}=K z_0^2 \Phi_c^4 \left( \ln \left( \frac{\Phi_c^2}{\Phi_v^2} \right)
- \half \right),
\EE
where
\BE
\label{K}
K = \frac{\eta}{8 \pi^2} \left[
\frac{(1+8\eta)(1-8\eta+32\eta^2+256\eta^3)}{(1+4\eta)^2} \right].
\EE

\subsection{Discussion}

  The constraint (\ref{re2}) arises from the requirement that the divergent
$I_{-1}$ terms in $\bar V_G$ cancel.  The equivalent constraint in pure O(2)
$\LPH$ analysis \cite{ON,brihaye} would fix the coefficient $\eta$ to be the
positive root of the numerator factor, $(1-8\eta-64\eta^2)$, which is
\BE
\label{eta0}
\eta_0 \equiv \frac{1}{4(1+\sqrt{5})} = 0.0773.
\EE
Here, however, one has instead a relationship between the two coupling
coefficients, which is shown in Fig. 1.  It is easily established that only
the region
between $\eta=0$ and $\eta = \eta_0$ is physically relevant.  This is because
(i) $\gamma$, being proportional to $e_B^2$, must be positive; and (ii) the
vector mass-squared $\bar\Delta^2$ must be positive, which precludes $\eta^2$
from being larger than $1/32$ (see Eq. (\ref{re1})).  From the figure we see
that there is a ``perturbative region'' in which both $\eta$ and $\gamma$
are small, with $\gamma^2 \approx (8/3)\eta$.  This corresponds to
$e^4 \sim {\cal O}(\lambda)$, as in Coleman and Weinberg (CW) \cite{cole}.
However, as $\eta$ increases, $\gamma^2$ reaches a maximum and then starts to
decrease, going to zero at $\eta = \eta_0$.  This extreme case corresponds
to a free vector theory completely decoupled from a self-interacting $\LPH$
theory.

  The vector-boson and Higgs masses come directly from Eq. (\ref{re1}),
evaluated at $\Phi_c=\Phi_v$.  Their ratio is given by:
\BE
\label{MHMV}
\frac{M_H^2}{M_V^2} =
\frac{\bar\OC_v^2}{\bar\Delta_v^2}
= \frac{8 \eta ( 1+ 4\eta)}{\gamma (1-32 \eta^2)},
\EE
which is just a function of $\eta$, since $\gamma$ is determined
by the constraint (\ref{re2}).  The mass-squared ratio is plotted in Fig. 2.
In the ``perturbative regime'' the Higgs is much lighter than the vector boson,
by a factor of $3 \gamma$, which is ${\cal O}(e^2)$ as in CW.  However, for
most of the range of $\eta$ the Higgs has a mass comparable to the vector.
When $\eta$ becomes close to $\eta_0$ the Higgs can be much heavier than
the vector.

  The other mass parameter, $\bar \OL^2$, does not have a direct physical
meaning.  It corresponds to the mass of the transverse scalar field, which
is, approximately, the Goldstone field.  In the covariant-gauge Higgs
mechanism \cite{higgs} the Goldstone field is an unphysical degree of freedom.
As discussed in Sect. 6, the fact that $\bar \OL^2$ is non-zero is due to
our approximation being unable to fully respect the O(2) symmetry.
We can therefore be pleased by the fact that $\OL^2$ is small (dashed line
in Fig. 2).

\subsection{Renormalization: Part II}

  We initially assumed that the mass-scale in the $I_{-1}$ denominator of
$e_B^2$ was the same as the mass-scale $\mu$ in $\LB$ (see Eq. (\ref{autren2}).
If this is not so then, using (\ref{key2}), we can re-write $e_B^2$ as:
\BE
e_B^2 = \frac{\gamma}{I_{-1}(\mu)} +
\frac{\gamma_2}{(I_{-1}(\mu))^2} + \ldots ,
\EE
where $\gamma_2$ is a coefficient proportional to the logarithm of the ratio
of the two mass-scales.  The subleading $\gamma_2/(I_{-1}(\mu))^2$ term
leads to an extra contribution, proportional to $\Phi_c^2$, in the
infinitesimal part of $\bar \OC^2$.  Thus, when $\bar V_G$ is obtained by
integrating (\ref{dvdphi}), we obtain an extra finite contribution
proportional to $\Phi_c^4$ in Eq. (\ref{rgep}).  However, if we then
re-parametrize the GEP in terms of the vacuum value $\Phi_v$, we obtain
Eq. (\ref{vgv}) {\it unchanged}:  all the differences are absorbed into the
relationship of $\Phi_v$ to $\mu$ and $\gamma_2$.  Exactly the same argument
applies if the scale in the $I_{-1}$ factor of $Z_{\phi}$ is different from
that in $\LB$ \cite{auto2}.  [The argument also applies if one wants to insist
upon replacing the factors of 3 in the GEP, representing the number of
polarization states of a massive vector field, by $3-\epsilon$ in dimensional
regularization.]

  Note that, for $m_B^2=0$, the bare Lagrangian is characterized by just
two bare parameters; $\LB$ and $e_B$.  Thus, we expect the renormalized GEP
to be characterized by two parameters.  This is indeed the case, and in the
final form, (\ref{vgv}), these are $\eta$ and $\Phi_v$.  (We shall shortly
see that $z_0$ is fixed in terms of $\eta$ by Eq. (\ref{re3}) below.)  The
$\Phi_v$ parameter has dimensions of mass, and its appearance constitutes
the ``dimensional transmutation'' phenomenon \cite{cole}.   Originally, the
``autonomous'' renormalization conditions (\ref{autren1}) and (\ref{autren2})
introduced a superfluity of parameters; $\eta$, $\gamma$, $z_0$, and the scale
arguments of the $I_{-1}$ factors.  As just discussed, it does not matter
if all these mass-scales are different, since they are eventually
subsumed in a single scale, $\Phi_v$.  We saw earlier that $\gamma$ was fixed
in terms of $\eta$ by the constraint (\ref{re2}), required for the $I_{-1}$
divergences to cancel.  It remains to show how $z_0$ is determined, and we
turn to this topic next.

  The ``autonomous'' renormalization involves a wavefunction
renormalization constant $Z_{\phi} = z_0 I_{-1}(\mu)$.  The $\LPH$
analysis in Refs. \cite{auto2,ON} set $z_0=1$ arbitrarily (although the
possibility of further finite re-scalings of the field was considered).
However, as Ref. \cite{consoli} has pointed out, $z_0$ is actually fixed
uniquely by the following argument.  The bare and renormalized two-point
functions are related by
\BE
\label{gambr}
\Gamma_B^{(2)} = Z_{\phi}^{-1} \Gamma_R^{(2)}.
\EE
Let us consider this relation at zero momentum in the vacuum $\PC=\varphi_v$.
$\Gamma_B^{(2)}$ is then given by the second derivative of the effective
potential, with respect to the bare field, at $\PC=\varphi_v$.  This is
easily calculated from (\ref{vgv}):
\BE
\left. \frac {d^2 \bar V_G}{d \varphi_c^2}\right|_{\PC=\varphi_v} =
\left. \frac {1}{Z_{\phi}} \frac {d^2 \bar V_G}{d \Phi_c^2}
\right|_{\Phi_c=\Phi_v} =
\frac {1}{Z_{\phi}}  8 K z_0^2 \Phi_v^2 .
\EE
The renormalized (Euclidean) two-point function (i.e., inverse propagator),
$\Gamma_R^{(2)}$, is just $p^2 + \bar\OC^2$ in the Gaussian approximation.
At zero momentum and at $\PC=\varphi_v$ it therefore becomes the
physical Higgs mass squared $M_H^2 = \bar\OC^2_v = 8 \eta z_0 \Phi_v^2$.
Hence, Eq. (\ref{gambr}) gives
\BE
\label{zm0}
\frac {1}{Z_{\phi}} 8 K z_0^2 \Phi_v^2
= \frac {1}{Z_{\phi}} 8 \eta z_0 \Phi_v^2,
\EE
which implies
\BE
\label {re3}
z_0(m_B \! = \! 0) = \frac{\eta}{K} = 8 \pi^2 \left[ \frac{(1+4 \eta)^2}
{(1 +8 \eta)( 1- 8\eta + 32 \eta^2 + 256 \eta^3)} \right].
\EE
The factor in square brackets varies between 1 and  1.536 for $\eta$
between 0 and $\eta_0$.  (See Fig. 3.)

  Finally, we remove our initial simplifying assumption that the bare mass
vanishes identically.  A finite bare mass would spoil the cancellation of
$I_{-1}$ divergences, but an infinitesimal bare mass,
\BE
m_B^2 = m_0^2/I_{-1}(\mu),
\EE
is allowed (Cf. Ref. \cite{auto2} with $I_0(0)=0$).  This produces an extra,
$\Phi_c$-independent, contribution to the $1/I_{-1}(\mu)$ part of $\bar \OC^2$.
Thus, when we integrate (\ref{dvdphi}), we obtain an extra finite contribution,
proportional to $\Phi_c^2$, in the GEP, Eq. (\ref{rgep}).  The result,
conveniently re-parametrized by $\Phi_v$ and a new parameter $m^2$
(trivially related to $m_0^2$), takes the form:
\BE
\label{vgv2}
\bar V_{G}=K z_0^2 \Phi_c^4 \left( \ln \left( \frac{\Phi_c^2}{\Phi_v^2} \right)
- \half \right) +
\half m^2 z_0
\Phi_c^2 \left( 1 - \half \frac{\Phi_c^2}{\Phi_v^2} \right) .
\EE
As before, $K$ is given by (\ref{K}) and $\Phi_v$ corresponds to the position
of the minimum of $\bar V_G$ \cite{fnssb}.

  Nothing else is affected except the determination of $z_0$.  The second
derivative of the GEP at the vacuum is now given by:
\BE
\left. \frac {d^2 \bar V_G}{d \varphi_c^2}\right|_{\PC=\varphi_v} =
\left. \frac {1}{Z_{\phi}} \frac {d^2 \bar V_G}{d \Phi_c^2}
\right|_{\Phi_c=\Phi_v} =
\frac {1}{Z_{\phi}} \left( 8 K z_0^2 \Phi_v^2 - 2 m^2 z_0 \right) ,
\EE
which replaces the left-hand side of Eq. (\ref{zm0}), so that we obtain
\BE
\label{z0}
z_0 = \frac{1}{K} \left( \eta  +
\frac{1}{4} \frac{m^2}{\Phi_v^2} \right).
\EE
Note that $m$ is not a particle mass.  In the symmetric vacuum all the
particles would be massless, for any $m^2$.  In the SSB vacuum the particle
masses are affected by $m^2$ only through its effect on $z_0$.

%
%

\section{Summary, Comparison to 1LEP, and Implications for the Higgs Mass}
\setcounter{equation}{0}

    We have calculated, with three different formalisms, the GEP of the
U(1) Higgs model.  The unrenormalized result, in a general covariant gauge,
is given at the end of Sect. 3.  In the optimal gauge, $\xi=0$, the result
is given in Sect. 6.

    To renormalize the GEP we postulated the infinitesimal forms
$\LB=\eta/I_{-1}$, $e_B^2=\gamma/I_{-1}$, $m_B^2=m_0^2/I_{-1}$ for the bare
parameters, and an infinite re-scaling of the classical field,
$\PC^2=z_0 I_{-1} \Phi_c^2$, where $I_{-1}$ is a log-divergent integral.
The cancellation of $I_{-1}$ divergences in the GEP gave the constraint
\BE
\gamma^2 = \frac{8 \eta}{3} \frac{(1 - 8 \eta - 64 \eta^2)}{(1-32 \eta^2)}.
\EE
(See Fig. 1.)  The vector and Higgs masses were found to be given by
\BA
M_V^2 & = & \gamma \frac{1-32\eta^2}{1+4\eta} z_0 \Phi_v^2,
\\
\label{higg}
M_H^2 & = & 8 \eta z_0 \Phi_v^2.
\EA
(See Fig. 2.)  The $z_0$ factor in the $\PC^2$ re-scaling was obtained in
Eq. (\ref{z0}):  in the $m_B=0$ case it varies between $8 \pi^2$ and
$(8 \pi^2)\times(1.536)$ (see (\ref{re3}) and Fig. 3).  The renormalized GEP,
Eq. (\ref{vgv2}) is a sum of $\Phi_c^4 \ln \Phi_c^2$, $\Phi_c^4$, and
$\Phi_c^2$ terms.

     At the unrenormalized level, we can recover the 1LEP simply by
discarding all the $I_0^2$ terms in Eq. (\ref{result0}), since each $I_0$
and $I_1$ is really accompanied by an $\hbar$ factor.  Consequently, the
optimization equations, (\ref{deq0} -- \ref{oeq0}), would be reduced to the
classical expressions,
$\Delta^2_c = e_B^2 \PC^2$,
$\Omega^2_c = m_B^2 + 12 \LB \PC^2$, \hspace*{2.5mm}
$\omega^2_c = m_B^2 +  4 \LB \PC^2$,
and Eq. (\ref{vbar}) would reduce to the familiar (unrenormalized)
1-loop result \cite{cole,jackiw}:
\BE
V_{1l} = \half m_B^2 \PC^2 + \LB \PC^4 +
3 I_1(\Delta_c) + I_1(\OC_c) + I_1(\OL_c).
\EE

    Conventionally, the 1LEP is renormalized in a perturbative fashion,
with $\LR = \LB (1 + {\cal O}(\LB \hbar I_{-1}) + \ldots )$, etc..  However,
it has been realized recently \cite{consoli,ims} that the 1LEP can also be
renormalized in an ``autonomous'' fashion.  The analysis exactly parallels
the GEP case, and can be made even simpler by directly using (\ref{i1key})
for $I_1$ \cite{ims}.  In the 1-loop case the constraint needed to cancel
the $I_{-1}$ divergences is:
\BE
\tilde\gamma^2 = \frac{8}{3}\tilde\eta(1-20\tilde\eta),
\EE
with tilde's distinguishing the 1-loop quantities from their GEP counterparts.
The vector and Higgs masses are given by
\BA
M_V^2 & = & \tilde\gamma \tilde{z}_0 \Phi_c^2,
\\
\label{hig1l}
M_H^2 & = & 12 \tilde\eta \tilde{z}_0 \Phi_c^2.
\EA
The $\tilde{z}_0$ factor in the massless case is $12 \pi^2$ (so one can regard
$12 \pi^2 \tilde{\gamma}$ as the renormalized $e^2$).  The renormalized
1LEP emerges (modulo the qualifications mentioned below) as:
\BE
\label{v1lren}
V_{1l} = 3 f(\Delta_c^2) + f(\OC_c^2) + f(\OL_c^2),
\EE
in terms of the function $f$ defined in Eq. (\ref{f}), and so is a mixture
of $\Phi_c^4 \ln \Phi_c^2$ and $\Phi_c^4$ terms.
[Actually, this result assumes $m_B=0$, and that all the $I_{-1}$
factors have the same scale $\mu$.  These assumptions are easily removed,
as discussed in ``Part II'' of the GEP analysis (Sect. 7.4):  one
simply gets additional $\Phi_c^2$ and $\Phi_c^4$ terms with free-parameter
coefficients.  The final result can again be parametrized in the form
(\ref{vgv2}).]

    Clearly, the autonomously renormalized 1LEP and GEP results have much
in common.  The 1LEP constraint equation and $M_H^2/M_V^2$ ratio are plotted
in Figs. 4 and 5.  Qualitatively, these closely resemble the GEP results
in Figs. 1 and 2.  In the 1-loop case the maximum $\tilde \eta$ is
$1/20 = 0.05$, rather than $\eta_0 = 0.077$, but a rescaling of the $\eta$
and $\gamma$ axes almost entirely absorbs the differences between the 1LEP
and GEP results.  The form of the renormalized potentials is also remarkably
similar, both when we compare (\ref{v1lren}) with (\ref{rgep}), and when we
note that they share the same final form (\ref{vgv2}).  In the 1-loop case
the coefficient $\tilde K$ would be given by $\tilde \eta/(8 \pi^2)$ instead
of by Eq. (\ref{K}).  These differ by the same factor that occurs in the
GEP's $z_0$; a factor that lies between 1 and 1.536.

    To see the implications for phenomenology, we can consider
$\Phi_v = (\sqrt{2}G_F)^{-1/2} = 246$ GeV, and $M_V \sim 90$ GeV.   This
implies a very small $\gamma$, and hence we must either be in the
perturbative regime where both $\eta$ and $\gamma$ are small, or near to the
maximum allowed $\eta$.  The former case gives a light Higgs, as in CW
\cite{cole}:  The latter case gives a Higgs that is much heavier than the
vector boson.   In fact, the Higgs mass would be almost exactly that of a
pure O(2) $\LPH$ theory whose $\Phi_v$ was 246 GeV.  For definiteness let us
assume the attractive possibility that the bare mass is zero \cite{cole}.
{}From the 1-loop result (\ref{hig1l}), with $\tilde \eta = 1/20$,
$\tilde z_0 = 12 \pi^2$, we would obtain $M_H = 2.07$ TeV.  From the GEP
result (\ref{higg}), with $\eta = \eta_0 = 0.0773$,
$z_0 = (8\pi^2) \times (1.533)$, we would obtain $M_H = 2.13$ TeV.  These
results agree remarkably well.

    Of course, these results are for the U(1) Higgs model, not the actual
SU(2)$\times$U(1) theory.  However, the 1-loop analysis is easily extended
to that theory \cite{ims}, and yields $M_H = 1.89$ TeV.  The GEP calculation
for SU(2)$\times$U(1) is a more difficult matter.  However, it is clear that
the phenomenological result would be essentially governed by the scalar sector,
which is an O(4), rather than an O(2), $\LPH$ theory.  The GEP results for the
O$(N)$ case can be obtained from Ref. \cite{ON}, supplemented by a quick
calculation of the proper $z_0$ factor, as explained in Sect. 7.4.  For
zero bare mass, this gives:
\BE
z_0[{\rm O}(N) \, \LPH ] = \frac{\pi^2}{2 \eta_0}
\frac{(1+4 \eta_0)}{(1-4 \eta_0)},
\EE
where $\eta_0$ in the O$(N)$ case is
\BE
\eta_0 = \frac{1}{4(1+ \sqrt{N+3})}.
\EE
The Higgs mass is again given by the form (\ref{higg}).  [If the bare mass
is non-zero, then the result is affected only by an ${\cal O}(m^2/\Phi_v^2)$
correction to $z_0$.]  The O(4) case gives us a GEP prediction for the
Standard-Model Higgs mass:
\[
M_H = 2.05 \,{\rm TeV}.
\]

Consoli {\it et al.} \cite{consoli} argue that the nonperturbative
renormalization used here implies a vanishing renormalized scalar
self-coupling (see also \cite{polonyi}).  That would drastically suppress
the Higgs-to-longitudinal-$W,Z$ couplings, leaving the Higgs with a relatively
narrow width.  The phenomenology of such a Higgs deserves urgent attention.

\vspace*{3mm}
\hspace*{-\parindent}{\bf Acknowledgements}

We would like to thank Maurizio Consoli, Jos\'e Latorre,
Anna Okopi\'nska, and Rolf Tarrach for very valuable discussions.
\\
This work was supported in part by the U.S. Department of Energy under
Contract No. DE-AS05-76ER05096.


\begin{thebibliography} {99}
\bibitem{fnte} The factor of 4 in the $\LB$ term is included so that the
O(2) $\LPH$ potential in Eq. (\ref{lscalar}) has the exactly the same
form as in Ref. \cite{ON}.
\bibitem{fnssb} This parametrization is only appropriate if an SSB minimum
exists, which need not be the case if $m_0^2$ is sufficiently large and
positive.
\end{thebibliography}
\end{document}